\documentstyle[seceq,wrapfig,epsf]{ptptex}
\newcommand\beq{\begin{equation}}
\newcommand\eeq{\end{equation}}
\newcommand\st{\sin\theta}
\newcommand\ct{\cos\theta}
\newcommand\esc{{e \over{\st \ct}}}
\newcommand\half{{1\over{2}}}
\newcommand{\NPB}[1]{{\it Nucl. Phys.}\ {\bf B{#1}}}
\newcommand{\PLB}[1]{{\it Phys. Lett.}\ {\bf B{#1}}}
\newcommand{\PRD}[1]{{\it Phys. Rev.}\ {\bf D{#1}}}
\renewcommand{\PRL}[1]{{\it Phys. Rev. Lett.}\ {\bf #1}}
\newcommand{\NCA}[1]{{\it Nuovo Cim.}\ {\bf {#1}A}}
\newcommand{\hc}{ {\rm h.c.} }
\newcommand{\ME}{ M_{ETC} }
\newcommand{\gE}{ g_{ETC} }

\notypesetlogo  %comment in if to eliminate PTPTeX logo

\title{%
Testing Extended Technicolor with $R_b$ \footnote{Talk given by
E.H.S. at the Yukawa International Seminar `95, Kyoto, August 21-15, 1995
and at the International Symposium on Heavy Flavor and Electroweak
Theory, Beijing, August 17-19, 1995. BU-HEP-95-27. hep-ph/9509392.} }

\author{%
Elizabeth H. {\sc Simmons}, {\sc R.S. Chivukula}, and J. {\sc
Terning}\footnote{E-mail: simmons@bu.edu, sekhar@bu.edu,
terning@calvin.bu.edu}
}

\inst{%
Physics Department, Boston University, 590 Commonwealth Ave., Boston
MA  02215
}

\recdate{%
\today
}

\abst{%
We review the connection between $m_t$ and
the $Zb\bar b$ vertex in ETC models and demonstrate the power of the
resulting experimental constraint on models with weak-singlet ETC bosons.
Some efforts to bring ETC models into agreement with experimental
data on the $Zb\bar b$ vertex are mentioned, and the most promising
one (non-commuting ETC) is discussed in detail.
}

\begin{document}

\maketitle

\section{Introduction}

Two outstanding questions in particle theory are the cause of
electroweak symmetry breaking and the origin of the masses and mixings
of the fermions.  Because theories that use light, weakly-coupled
scalar bosons to answer these questions suffer from the hierarchy and
triviality problems, it is interesting to consider the possibility that
electroweak symmetry breaking arises from strong dynamics at scales of
order 1 TeV.  This talk focuses on extended\cite{ETC}
technicolor\cite{tc} (ETC) models,
in which both the masses of the weak gauge bosons and those of the
fermions arise from gauge dynamics.

In extended technicolor models, the large mass of the top quark
generally arises from ETC dynamics at relatively
low energy scales.  Since the magnitude of the
Cabibbo-Kobayashi-Maskawa matrix element $\vert V_{tb}\vert$ is nearly
unity, $SU(2)_W$ gauge invariance insures that ETC bosons coupling to
the left-handed top quark couple with equal strength to the
left-handed bottom quark.   In particular, the
ETC dynamics
which generate the top quark's mass also couple to the left-handed
bottom quark thereby affecting the $Zb\bar b$ vertex.  This has been
shown\cite{zbbone} to provide
a powerful experimental constraint on extended technicolor models --
particularly on those models in which the ETC gauge group commutes
with $SU(2)_W$.

This talk begins by reviewing the connection between the top quark
mass and the $Zb\bar b$ vertex in ETC models.  Next, the power of the
resulting experimental constraint on ETC models with weak-singlet ETC
bosons is demonstrated.  Several recent attempts
\cite{walkrb,strongETC,comp,evans,yoshi,NCETC,NCETCtwo}
to bring ETC
models into agreement with experimental data on the $Zb\bar b$ vertex
are mentioned, and the most promising one (non-commuting ETC) is
discussed.

\section{From $m_t$ To A Signal of ETC Dynamics}

 Consider a model in which $m_t$
is generated by the exchange of  a
weak-singlet ETC gauge boson of mass $M_{ETC}$ coupling  with
strength $\gE$ to the current
\begin{equation}
{\xi} {\bar\psi^i}_L \gamma^\mu T_L^{ik}
+ {1\over\xi} {\bar t_R} \gamma^\mu U_R^k\ ,\ \ \ \ \ \ {\rm where}\ \
\psi_L\ \equiv\ \pmatrix{t \cr b \cr}_L\ \
T_L\ \equiv\ \pmatrix{U \cr D \cr}_L
\label{tmasscur}
\end{equation}
where $U$ and $D$ are technifermions, $i$ and $k$ are weak and technicolor
indices, and $\xi$ is an ETC Clebsch expected
to be of order one.  At energies below $\ME$, ETC  gauge boson exchange
may be approximated by local four-fermion operators.   For example, $m_t$
arises from an operator coupling the  left- and right-handed currents
in Eq. (\ref{tmasscur})
\beq
   - {\gE^2 \over  \ME^2}  \left({\bar\psi}_L^i \gamma^\mu
T_L^{iw}\right) \left( {\bar U^w}_R \gamma_\mu t_R \right) + \hc\ .
\label{topff}
\eeq
When this is Fierzed into a product of technicolor-singlet densities, it
generates a mass for the top quark when the technifermions' chiral
symmetries break. We can use the rules of naive  dimensional analysis
\cite{dimanal} to estimate the size of $m_t$ generated by
Eq. (\ref{topff}). Assuming, for simplicity, that there is only a
doublet of technifermions and that technicolor respects an
 $SU(2)_L \times SU(2)_R$ chiral symmetry (so that the technipion
decay constant, $F$, is $v= 246$ GeV) we have
\beq
   m_t\ = {\gE^2 \over \ME^2}
   \langle{\bar U}U\rangle\ \approx\ {\gE^2 \over \ME^2} (4\pi v^3)\ .
\label{topmass}
\eeq

The ETC boson responsible for
producing $m_t$ also affects the $Zb\bar b$ vertex\cite{zbbone}.
Consider the four-fermion
operator arising from the left-handed current in Eq.(\ref{tmasscur})
-- the part containing $b$ quarks.
\beq
  -\xi^2 {\gE^2\over\ME^2} \left({\bar \psi^i}_L \gamma^\mu
  T^{iw}_L \right) \left({\bar T^{jw}}_L \gamma_\mu \psi^j_L \right)\ .
\label{unfierz}
\eeq
When Fierzed into a product of technicolor singlet currents, this
includes
\beq
  -{\xi^2\over 2} {\gE^2\over
  \ME^2}\left({\bar\psi}_L\gamma^\mu\tau^a\psi_L \right) \left({\bar T}_L
  \gamma_\mu \tau^a T_L \right)\ ,
\label{fourferm}
\eeq
where the $\tau^a$ are weak isospin Pauli matrices.

Adopting an effective chiral Lagrangian description appropriate below
the technicolor chiral symmetry breaking scale, we may replace the
technifermion current by a sigma-model current \cite{howardbook}:
\beq
 \left({\bar T}_L \gamma_\mu \tau^a T_L \right) =
	{v^2 \over 2}Tr\left(\Sigma^\dagger\tau^a iD_\mu\Sigma\right)\
,
\eeq
where $\Sigma = \exp{(2i{\tilde\pi}/v)}$ transforms as
$\Sigma \rightarrow L\Sigma R^\dagger$ under $SU(2)_L \times SU(2)_R$,
and
\beq
D_\mu \Sigma = \partial_\mu\Sigma
\ +\ i\esc Z_\mu\left({\tau_3\over 2}\Sigma -
\st^2[Q,\Sigma]\right)\ +\ ...
\eeq
In unitary gauge $\Sigma=1$ this is seen to alter the $Z$-boson's
tree-level coupling
to left-handed bottom quarks $g_L = \esc(-\half + {1\over 3}\st^2)$ by
\begin{eqnarray}
\delta g_L^{ETC} &=& -{\xi^2 \over 2} {\gE^2 v^2\over\ME^2} \esc(I_3)
\label{ta}\\
	&=& {1\over 4} {\xi^2} {m_t\over{4\pi v}}
\cdot \esc \label{tb}
\end{eqnarray}
Here eq. (\ref{tb}) follows from applying eq. (\ref{topmass}) to eq.
(\ref{ta}).

\section{Comparison with the Standard Model and LEP data}

To show that $\delta g_L$ provides a test of ETC dynamics, we must
relate it to a shift in the value of an experimental observable, and
compare that shift both to radiative corrections in the
standard model and to the available experimental precision.

Because ETC gives a direct correction to the $Zb\bar b$ vertex, we
need an observable that is particularly sensitive to direct, rather
than oblique\cite{ST}, effects.  A natural choice is the ratio of $Z$
decay widths
\beq
R_b \equiv {\Gamma(Z\to b\bar b) \over {\Gamma(Z \to {\rm hadrons})}}
\eeq
because both the oblique and QCD corrections largely cancel in this ratio.
As
\beq
{\delta \Gamma_b \over {\Gamma_b}} = 2 { g_L \delta g_L \over {g_L^2 +
g_R^2}} \approx  -6.5\%\ \xi^2\ \left({m_t\over {175 {\rm GeV}}}\right)
\eeq
we find
\beq
{\delta R_b \over R_b} \approx -5.1\% \xi^2 \left({m_t
\over 175{\rm GeV}} \right).
\label{rbb}
\eeq
The one-loop $Zb\bar b$ vertex correction in the standard model, which
is largely due to exchange of longitudinal $W$ bosons, lies in the
range $[-0.5\% ... -2.0\%]$\cite{tcalc} for 100 GeV $\geq m_t \geq$
200 GeV.  The ETC-induced correction (\ref{rbb}) is larger and in the
same direction.  Furthermore, because ETC models include longitudinal
$W$ bosons, the full shift in $R_b$ in an ETC model is the sum of the
$W$-exchange and ETC contributions.

The LEP experiments now have sufficient precision to detect such large
shifts in $R_b$.  The experimental value of $R_b = 0.2202\pm 0.0020$
actually lies {\it above} the 1-loop standard model value of
$R_b = 0.2155$ \cite{langacker,blondel}. This implies that any
contribution from non-standard physics is positive: $\left[\delta R_b/
R_b\right]_{new} \approx + 2.2\%$, thereby excluding ETC models in
which the ETC and weak gauge groups commute.

\section{Interlude}

Having demonstrated that measurements of $R_b$ can exclude a
significant class of simple ETC models, we should check how
more realistic models fare.   Accordingly, we briefly review the
impact of certain features of recent ETC models on the value of
 $R_b$.

A slowly-running (`walking') technicolor beta-function is often
included in ETC models in order to provide the light fermions with
realistically large masses, while avoiding excessive flavor-changing
neutral currents\cite{walktc}.  Because a walking beta function
enhances the size
of the technifermion condensate $\langle \bar T T \rangle$, it leads
to larger fermion masses for a given ETC scale, $M_{ETC}$.  Enhancing
 $m_t$ relative to $M_{ETC}$ reduces the size of $\delta g_L$.
However, it is has been shown \cite{walkrb} that the shift in $R_b$
generally remains large enough to be visible at LEP.

It is possible to build ETC models in which the ETC coupling itself
becomes strong before the scale $M_{ETC}$ and plays a significant role
in electroweak symmetry breaking \cite{strongETC}.  The spectrum of
strongly-coupled ETC models include light composite scalars with
Yukawa couplings to ordinary fermions and technifermions \cite{comp} .
Exchange of the composite scalars produces corrections to $R_b$ that
are small enough to leave $R_b$ in agreement with experiment
\cite{evans}.  The disadvantage of this approach is the need to
fine-tune the ETC coupling close to the critical value.

ETC models also generally include `diagonal' techni-neutral ETC
bosons.  The effect of these gauge bosons on $R_b$ is discussed at
length in Ref. \citen{yoshi}.  Suffice it to say that while exchange of
the diagonal ETC bosons does tend to raise $R_b$, this effect is
significant only when the model includes large isospin violation --
leading to conflict with the measured value of the oblique parameter
$T$.

Finally, we should recall that our analysis explicitly assumed that
the weak and ETC gauge groups commute.  More recent work
\cite{NCETC,NCETCtwo} indicates that relaxing that assumption can
lead to models with experimentally acceptable values of $R_b$.
The remainder of this talk will therefore focus on `non-commuting'
extended technicolor models.

\section{Non-commuting ETC Models}

We begin by describing the symmetry-breaking pattern that enables
non-commuting ETC models to include both a heavy top quark and
approximate Cabibbo universality \cite{NCETC}.  A heavy top quark must
receive its
mass from ETC dynamics at low energy scales; if the ETC bosons
responsible for $m_t$ are weak-charged, the weak group $SU(2)_{heavy}$
under which $(t,b)_L$ is a doublet must be embedded in the low-scale
ETC group.  Conversely, the light quarks and leptons cannot be charged
under the low-scale ETC group lest they also receive large
contributions to their masses; hence the weak $SU(2)_{light}$ group
for the light quarks and leptons must be distinct from
$SU(2)_{heavy}$.  To approximately preserve low-energy Cabibbo
universality the two weak $SU(2)$'s must break to their diagonal
subgroup before technicolor dynamically breaks the remaining
electroweak symmetry.  The resulting
symmetry-breaking pattern is
%\footnote{The hypercharge group, $U(1)_Y$,
%is embedded partly in the ETC group, so that $U(1)' \neq U(1)_Y$.}
:
\begin{eqnarray}
ETC & \times& SU(2)_{light} \times U(1)' \nonumber\\
&\downarrow&\ \ \ \ \ f \nonumber \\
TC \times SU(2)_{heavy} & \times& SU(2)_{light} \times U(1)_Y  \nonumber\\
&\downarrow&\ \ \ \ \ u \\
TC & \times& SU(2)_W \times U(1)_Y \nonumber \\
&\downarrow&\ \ \ \ \ v \nonumber \\
TC & \times& U(1)_{EM}, \nonumber
\end{eqnarray}
\noindent{where $ETC$ and $TC$ stand, respectively, for the extended
technicolor and technicolor gauge groups, while $f$, $u$, and
$v = 246$ GeV are the expectation values of the order parameters for the
three different symmetry breakings ({\it i.e.} the analogs of $F_\pi$
for chiral symmetry breaking in QCD).  Note that, since we are
interested in the physics associated with top-quark mass generation,
only $t_L$, $b_L$ and $t_R$ need transform non-trivially under $ETC$.
But to ensure anomaly cancelation, it is more economical to assume
that the entire third generation has the same non-commuting ETC
interactions.  Thus we take $(t,b)_L$ and $(\nu_\tau,\tau)$ to be
doublets under $SU(2)_{heavy}$ but singlets under $SU(2)_{light}$,
while all other left-handed ordinary fermions have the opposite
$SU(2)$ assignment.}

Once again, the dynamics responsible for generating the top quark's
mass contributes to $R_b$.  This time the ETC gauge boson involved
transforms as a weak doublet coupling to
\beq
\xi\bar\psi_L \gamma^\mu U_L +
{1\over \xi}\bar t_R \gamma^\mu T_R
\eeq
where $\psi_L \equiv (t,b)_L$ and $T_R \equiv (U,D)_R$, are doublets
under $SU(2)_{heavy}$ while $U_L$ is an $SU(2)_{heavy}$ singlet.
Therefore, the four-fermion operator affecting the $Zb\bar b$ operator
is (after Fierzing)
\beq
- {2\xi^2\over f^2} \left( \bar\psi_L \gamma^\mu \psi_L
\right) \left( \bar U_L \gamma_\mu U_L \right) .
\label{a:2}
\eeq

Note that, since $\bar\psi_L \gamma^\mu
U_L$ and $\bar t_R \gamma^\mu T_R$ must transform in the same way, if
 $\psi_L$ is a 2 of $SU(2)_{heavy}$ then $T_R$ is a 2* instead.
Then in the effective chiral Lagrangian description, we find
%\beq
%D_\mu T_R = \partial_\mu T_R + i \esc Z_\mu ({1 \over
%2}(-\tau_3^*) -s^2 Q) T_R + . . .
%\eeq
%so that
\beq
D_\mu \Sigma = \partial_\mu \Sigma - {{i e}\over{\st \ct}} Z_\mu
\left({1\over 2} \Sigma \tau_3^*  + \sin^2\theta[\Sigma,Q]\right) + ...
\eeq
and in unitary gauge ($\Sigma = 1$) we have
\beq
\bar U_L \gamma_\mu U_L = - \esc {v^2 \over
4} Z_\mu.
\eeq
The effect of operator (\ref{a:2}) on the coupling of $b_L$ to the $Z$
is therefore
\beq
\delta g_L = -\esc {\xi^2 v^2 \over { 2 f^2}} \approx -
{\xi^2 \over 4} \esc {m_t \over {4 \pi v}} .
\label{a:1}
\eeq
 Since the tree-level $Z b_L
\bar b_L$ coupling is also negative, the ETC-induced change tends to
{\bf increase} the coupling -- and thereby increase $R_b$.  We find
that Eq. (\ref{a:1}) results in a change to $R_b$ of\cite{NCETCtwo}
\beq
{\delta R_b \over R_b} \approx +5.1\% \xi^2 \left({m_t
\over 175{\rm GeV}} \right).
\eeq
The change is similar in size to what was obtained in the
commuting ETC models, but is opposite in sign.

But that is not the full story of $R_b$ in non-commuting ETC.  Recall
that there are two sets of weak gauge bosons which mix at the scale
$u$.  Of the resulting mass eigenstates, one set is heavy and couples
mainly to the third-generation fermions while the other set is {\it
nearly} identical to the $W$ and $Z$ of the standard model.  That
`nearly' is important: it leads to a shift in the light $Z$'s coupling
to the $b$ of order\cite{NCETCtwo}
\beq
\delta g_L = {e \over{2 \st \ct}} {g_{ETC}^2 v^2\over u^2} \sin^2\alpha
\eeq
where $\tan\alpha = g_{light}/g_{heavy}$ is the ratio of the $SU(2)$
gauge couplings.  The couplings of the light $Z$ to other fermions are
similarly affected.  When this is included, mixing alters
$R_b$ by
\beq
{\delta R_b \over R_b} \approx -5.1\%
\sin^2\alpha {f^2\over u^2} \left({m_t
 \over 175{\rm GeV}} \right).
\eeq
The two effects on $R_b$ in non-commuting ETC models
are of similar size and opposite sign, and their precise values are
model-dependent.  Thus, non-commuting ETC theories can
yield values of $R_b$ that are consistent with experiment\cite{NCETCtwo}.

Since $R_b$ alone cannot confirm or exclude non-commuting ETC, we
should apply a broader set of precision electroweak tests.  Before
doing this, we must describe the $SU(2)\times SU(2)$ symmetry breaking
sector in more detail.  The two simplest possibilities for the
$SU(2)_{heavy} \times SU(2)_{light}$ transformation properties of the
order parameters that produce the correct combination of mixing and
breaking of these gauge groups are:
\begin{eqnarray}
&\langle \varphi \rangle& \sim (2,1)_{1/2},\ \ \ \ \langle
\sigma\rangle \sim (2,2)_0 ~,\ \ \ \ \ \ \ ``{\rm heavy\ case}"\\
&\langle \varphi \rangle& \sim (1,2)_{1/2},\ \ \ \ \langle
\sigma\rangle \sim (2,2)_0 ~,\ \ \ \ \ \ \ ``{\rm light\ case}"~.
\end{eqnarray}
Here the order parameter $\langle\varphi\rangle$ is responsible for
breaking $SU(2)_L$ while $\langle\sigma\rangle$ mixes
$SU(2)_{heavy}$ with $SU(2)_{light}$.  We refer to these two
possibilities as ``heavy'' and ``light'' according to whether
$\langle\varphi\rangle$
transforms non-trivially under $SU(2)_{heavy}$ or $SU(2)_{light}$.

The heavy case, in which $\langle\varphi\rangle$ couples to the heavy
group, is the choice made in \cite{NCETC}, and corresponds to the case
in which the technifermion condensation responsible for providing mass
for the third generation of quarks and leptons is also responsible for
the bulk of electroweak symmetry breaking.  The light case, in which
$\langle\varphi\rangle$ couples to the light group, corresponds to the
opposite scenario in which different physics provides mass to the
third generation fermions and the weak gauge bosons.  While this light
case is counter-intuitive (after
all, the third generation is the heaviest!), it may provide a
resolution to the issue of how large isospin breaking can exist in the
fermion (and technifermion) mass spectrum without leaking into the $W$
and $Z$ masses.

We have performed a global fit for the parameters of the non-commuting
ETC model ($s^2$, $1/x \equiv v^2/u^2$, and the $\delta g$'s) to all
precision electroweak data: the $Z$ line shape, forward backward
asymmetries, $\tau$ polarization, and left-right asymmetry measured at
LEP and SLC; the $W$ mass measured at FNAL and UA2; the electron and
neutrino neutral current couplings determined by deep-inelastic
scattering; the degree of atomic parity violation measured in Cesium;
and the ratio of the decay widths of $\tau \to \mu \nu \bar\nu$ and
$\mu\to e \nu \bar
\nu$.  Details of the calculation are reported in \cite{NCETCtwo}.

Table \ref{Pred} compares the predictions of the standard model
and the non-commuting ETC model (for particular values of $1/x$ and
$s^2$) with the experimental values.  For $s^2$, we have chosen a
value of 0.97, at which the ETC gauge coupling is strong yet does not
break the technifermion chiral symmetries by itself\cite{NCETCtwo}.
For $1/x$, in the heavy case we show the best fit
value of $1/x =0.0027$ or equivalently $M_W^H= 9$ TeV.  The best fit
for $1/x$ in the light case lies in the unphysical region of negative
 $x$ but has large uncertainty:  $1/x = -0.17 \pm 0.75$.  For
illustration, we choose a value of $1/x$ from the large range of
values that give a good fit to the data; our choice, $1/x = 0.055$,
corresponds to $M_W^H = 2$ TeV.  We use\cite{alphasmall,langacker}
$\alpha_s(M_Z) = 0.115$ in these fits.

\begin{table}[htbp]
\begin{center}
\begin{tabular}{|c|l|l|l|l|}\hline\hline
Quantity & Experiment & SM & ${\rm ETC}_{\rm heavy}$ &
${\rm ETC}_{\rm light}$ \\\hline \hline
$\Gamma_Z$ & 2.4976 $\pm$ 0.0038 & 2.4923 & 2.4991 & 2.5006 \\
$R_e$ & 20.86 $\pm$ 0.07 & 20.73 & 20.84 & 20.82 \\
$R_\mu$ & 20.82 $\pm$ 0.06 & 20.73 & 20.84 & 20.82 \\
$R_\tau$ & 20.75 $\pm$ 0.07 & 20.73 & 20.74 & 20.73 \\
$\sigma_h$ & 41.49 $\pm$ 0.11 & 41.50 & 41.48 & 41.40 \\
$R_b$ & 0.2202 $\pm$ 0.0020 & 0.2155 & 0.2194 & 0.2188 \\
$A_{FB}^e$ & 0.0156 $\pm$ 0.0034 & 0.0160 & 0.0159 & 0.0160 \\
$A_{FB}^\mu$ & 0.0143 $\pm$ 0.0021 & 0.0160 & 0.0159 & 0.0160 \\
$A_{FB}^\tau$ & 0.0230 $\pm$ 0.0026 & 0.0160 & 0.0164 & 0.0164 \\
$A_{\tau}(P_\tau)$ & 0.143 $\pm$ 0.010 & 0.146 & 0.150 & 0.150 \\
$A_{e}(P_\tau)$ & 0.135 $\pm$ 0.011 & 0.146 & 0.146 & 0.146 \\
$A_{FB}^b$ & 0.0967 $\pm$ 0.0038 & 0.1026 & 0.1026 & 0.1030 \\
$A_{FB}^c$ & 0.0760 $\pm$ 0.0091 & 0.0730 & 0.0728 & 0.0730 \\
$A_{LR}$ & 0.1637 $\pm$ 0.0075 & 0.1460 & 0.1457 & 0.1460 \\
$M_W$ & 80.17 $\pm$ 0.18 & 80.34 & 80.34 & 80.34 \\
$M_W/M_Z$ & 0.8813 $\pm$ 0.0041 & 0.8810 & 0.8810 & 0.8810 \\
$g_L^2(\nu N \rightarrow \nu X)$ & 0.3003 $\pm$ 0.0039 & 0.3030 & 0.3026 &
0.3030  \\
$g_R^2(\nu N \rightarrow \nu X)$ & 0.0323 $\pm$ 0.0033 & 0.0300 & 0.0301 &
0.0300  \\
$g_{eA}(\nu e \rightarrow \nu e)$ & -0.503 $\pm$ 0.018 & -0.506 & -0.506 &
-0.506 \\
$g_{eV}(\nu e \rightarrow \nu e)$ & -0.025 $\pm$ 0.019 & -0.039 & -0.038 &
-0.039 \\
$Q_W(Cs)$ & -71.04 $\pm$ 1.81 & -72.78 &  -72.78 & -72.78 \\
$R_{\mu \tau}$ & 0.9970 $\pm$ 0.0073 & 1.0 & 0.9946 & 1.0  \\
\hline\hline
\end{tabular}
\end{center}
\vskip 0.1in
\caption{Experimental \protect\cite{langacker,blondel,taudec} and
predicted values of
electroweak observables for the standard model and non-commuting ETC
model (heavy and light cases) for $\alpha_s(M_Z)=0.115$, and $s^2=0.97$.
For the heavy case $1/x$ is allowed to assume the best-fit value of
0.0027; for the light case, $1/x$ is set to  $0.055$. The standard
model values
correspond to the best-fit values (with $m_t=173$ GeV, $m_{\rm Higgs}
= 300$ GeV) in \protect\cite{langacker}, corrected for the change in
$\alpha_s(M_Z)$, and the revised extraction \protect\cite{swartz} of
$\alpha_{em}(M_Z)$.}
\label{Pred}
\end{table}

Table \ref{lightFit115} illustrates how well non-commuting ETC models
fit the precision data. This table shows the fit to the standard
model for comparison; as a further benchmark we have included a fit to
purely oblique corrections (the $S$ and $T$ parameters) \cite{ST}.
The percentage quoted in the Table is the probability of obtaining a
 $\chi^2$ as large or larger than that obtained in the fit, for the
given number of degrees of freedom (df), assuming that the model is
correct.  Thus a small probability corresponds to a poor fit.  The
$SM+S,T$ fit demonstrates that merely having more parameters does not
ensure a better fit.

\begin{table}[htbp]
\begin{center}
\begin{tabular}{|l||c||c||c||c|}\hline\hline
Model &  $\chi^2$&df&$\chi^2/{\rm df}$  & probability \\ \hline\hline
SM & $33.8$ & $22$& $1.53$ & $5\%$ \\ \hline
SM+$S$,$T$  &$32.8$ & $20$ & $1.64$ & $4\%$ \\ \hline
${\rm ETC}_{\rm light}$  &$22.6$ & $20$ & $1.13$ & $31\%$ \\ \hline
${\rm ETC}_{\rm heavy}$  &$20.7$ & $19$ & $1.09$ & $36\%$ \\ \hline
\hline
\end{tabular}
\end{center}
\vskip 0.1in
\caption{The best fits for the standard model, beyond the standard model
allowing $S$  and $T$
to vary, and the non-commuting ETC model (heavy and light cases).  The
inputs are: $\alpha_s(M_Z)=0.115$, $s^2=0.97$ (for both ETC models), and
 $1/x=0.055$ (light case ETC). $\chi^2$ is the sum
of the squares of the difference between prediction and experiment,
divided by the error.}
\label{lightFit115}
\end{table}

{}From Tables \ref{Pred} and \ref{lightFit115}
we see that because non-commuting ETC models accommodate
changes in the $Z$ partial widths, they give a
significantly better fit to the experimental data than the standard
model does, even after taking into account that in the fitting
procedure the non-commuting ETC models have two extra parameters.
In particular non-commuting ETC predicts values for
$\Gamma_Z$, $R_e$, $R_\mu$, $R_\tau$,  and $R_b$ that are closer to
experiment than those predicted by the standard model.

For comparison we also performed the fits using\cite{langacker}
$\alpha_s(M_Z)=0.124$;
the quality of the fit is summarized in Table \ref{lightFit124}.
While the standard model fit improves for a
larger value of $\alpha_s(M_Z)$, the light case of the non-commuting
ETC model remains a better fit.

\begin{table}
\begin{center}
\begin{tabular}{|l||c||c||c||c|}\hline\hline
Model &  $\chi^2$&df&$\chi^2/{\rm df}$  & probability \\ \hline\hline
SM & $27.8$ & $21$& $1.33$ & $15\%$ \\ \hline
SM+$S$,$T$  &$27.7$ & $20$ & $1.38$ & $12\%$ \\ \hline
 ${\rm ETC}_{\rm light}$  &$25.0$ & $20$ & $1.25$ & $20\%$ \\ \hline
 ${\rm ETC}_{\rm heavy}$  &$22.3$ & $19$ & $1.17$ & $27\%$ \\ \hline
\hline
\end{tabular}
\end{center}
\vskip 0.1in
\caption{The best fits for the standard model, beyond the standard model
allowing $S$ and $T$
to vary, and the non-commuting ETC model.  The inputs are:
 $\alpha_s(M_Z)=0.124$, $s^2 = 0.97$, and for the light case $1/x=0.055$.}
\label{lightFit124}
\end{table}

%
%%%
%%%FIG-1 GOES HERE
%%%
\begin{figure}
\epsfxsize=4.5truein
\centerline{\epsffile{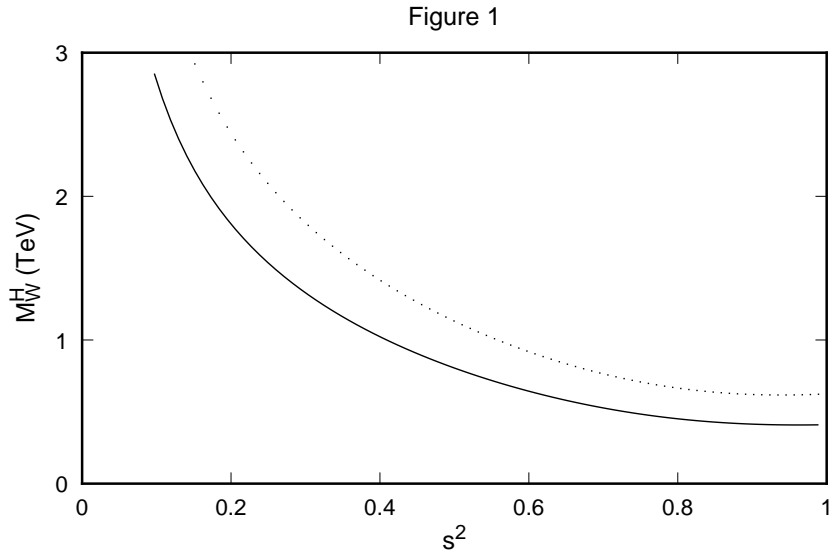}}
\vskip -3in
\caption{Figure 1. Lower bound on $M_W^H$ at 95\% c.l. (solid
line) and 68\% c.l. (dotted line) as a function
of $s^2$ for the light case (using $\alpha_s(M_Z) = 0.115$)}
\label{fig:1}
\end{figure}

\vskip -.2cm
As a bonus, the extra $W$ and $Z$ bosons can be
relatively light.  Figure 1 displays the 95\% and 68\% confidence
level lower bounds
(solid and dotted lines) on the heavy $W$ mass ($M^H_{W}$) for different
values of $s^2$ (with $\alpha_s(M_Z)=0.115$).  The plot was
created as follows: for each value of $s^2$ we fit to the three
independent parameters ($\delta g_L^b$, $\delta g_L^\tau = \delta
g_L^{\nu_\tau}$, and $1/x$); we then found the lower bound on $x$ and
translated it into a lower bound on the heavy $W$ mass.
Note that for $s^2 > 0.85$, the heavy $W$ gauge boson can be as light
as $400$ GeV. In the heavy case, similar work shows that the lowest
possible heavy $W$ mass at the 95\% confidence level  is $\approx 1.6$
TeV, for $0.7< s^2 <0.8$.

\section{Conclusions}

The $Z b\bar b$ vertex is sensitive to the dynamics that generates the
top quark mass.  As such, it provides an excellent test of extended
technicolor models. Measurements of $R_b$ at LEP have already excluded
ETC models in which the ETC and weak gauge groups commute.  Models in
which ETC gauge bosons carry weak charge can give experimentally
allowed values of $R_b$ because contributions to the $Zb\bar b$ vertex
from $Z Z'$ mixing are similar in magnitude and opposite in sign to
those from exchange of the ETC boson that generates the top quark's
mass.  These non-commuting ETC models can actually fit the full set of
precision electroweak data better than the standard model.

\medskip
This work was supported in part by NSF grants PHY-9057173 and
PHY-9501249, and by DOE grant
DE-FG02-91ER40676.

\end{document}